\documentclass[aps,twocolumn,superscriptaddress]{revtex4}
\usepackage{bm}
\usepackage{graphicx}
\usepackage{amsmath}

\usepackage{color}

\begin{document}

\title{Influence of sample preparation on the transformation of low-density to high-density amorphous ice: An explanation based on the Potential Energy Landscape}

\author{Nicolas Giovambattista}
\affiliation{Department of Physics, Brooklyn College of the City University of New York, Brooklyn, NY 11210 USA}
\affiliation{Ph.D. Programs in Chemistry and Physics, The Graduate Center of the City University of New York, New York, NY 10016}

\author{Francis W. Starr}

\affiliation{Department of Physics, Wesleyan University, Middletown, Connecticut 06459, United States}

\author{Peter H. Poole}
\affiliation{Department of Physics, St. Francis Xavier University,
Antigonish, NS, B2G 2W5, Canada}

\date{\today --- published as J. Chem. Phys. 147, 044501 (2017)}

\begin{abstract}
Experiments and computer simulations of the transformations of amorphous ices display different behavior depending on sample preparation methods, and on the rates of change of temperature and pressure to which samples are subjected.  In addition to these factors, simulation results also depend strongly on the chosen water model.
Using computer simulations of the ST2 water model, we study how the sharpness of the
 compression-induced transition from low-density amorphous ice (LDA) to high-density amorphous ice (HDA) 
is influenced by the preparation of LDA.
By studying LDA samples prepared using widely different procedures,
we find that the sharpness of the LDA-to-HDA transformation is correlated with
 the depth of the initial LDA sample in the potential energy landscape (PEL), 
as characterized by the inherent structure energy.
Our results show that the complex phenomenology of the amorphous ices reported in experiments 
and computer simulations 
can be understood and predicted in a unified way from a knowledge of the PEL of the system.
\end{abstract}

\maketitle

\section{Introduction}

The experimentally observed compression-induced transformation between low-density 
amorphous ice (LDA) and high-density amorphous ice (HDA) is remarkably sharp, and 
reminiscent of an equilibrium first-order phase 
transition~\cite{mishimaNature1985,aokiMishima,koza2003,susukiMishima2002,mishimaJPC1994,katrinRevers2008_2}.
However, the sharpness of the LDA-HDA transformation is sensitive to 
relaxation effects, compression rates, and sample 
preparation details~\cite{tulk2002,guthrie2003,klotz2005,thomasLDAs,finneyLoertingReview,katrinSnapshots,katrin2006,thomasT_HDAtoLDA,nelmes}.
 This variability has been cited as evidence to refute the association of phase-transition-like
 characteristics to the LDA-HDA transformation, and thus weaken the case in support of 
the closely related liquid-liquid phase transition (LLPT) hypothesis for 
supercooled water (see, e.g., Refs.~\cite{tulk2002,guthrie2003,klotz2005,commentToklotz,replyKlotz,tseKlug}).

Computer simulations of amorphous ice also show that the sharpness of the LDA-HDA
 transformation may vary with the compression rate considered (see, e.g., Refs.~\cite{martonak,tseKlein}).
  Simulations also give dramatically different results depending on the water model employed.
Simulations using the ST2 water model show a sharp LDA-HDA transformation for
 appropriate cooling and compression rates, while in
simulations of the SPC/E water model using the same rates, 
the LDA-HDA transformation is much more gradual~\cite{MySciRep,chiu1,yoPhaseDiag}.
A unified framework, based on statistical mechanics, 
that explains the complex phenomenology of amorphous ice, as 
observed in both experiments and computer simulations, is lacking
at the present time.

In this work, we perform extensive molecular dynamics simulations of water 
to study the pressure-induced LDA-HDA transformation using a controlled set of initial
 LDA samples. We use the ST2 water model~\cite{ST2model},
 which exhibits a well characterized LLPT that 
separates a low-density liquid (LDL) from a high-density liquid (HDL) phase in the 
supercooled region of the phase diagram under conditions where the liquid can be 
observed in (metastable) 
equilibrium~\cite{PooleNature92,geneLLCP2013,palmer,liuPablo,pooleSaika,ST2modelsLLCP,poole1993}.
As stated above, the amorphous solid form of ST2 water has also been shown
 to qualitatively reproduce the glass phenomenology of real water, 
including the LDA-HDA transformation, when subjected to appropriate cooling 
and compression rates~\cite{chiu1,chiu2,MySciRep,poole1993}.

The focus of the present work is the question of how the procedure used to prepare the initial sample of LDA affects the sharpness of the subsequent LDA-HDA transformation.  
As described below, we create initial LDA samples using several distinct methods, and then compress each until the system converts to HDA. We find that, 
depending on the LDA preparation method, the LDA-HDA transformation can be either smooth and gradual, or sharp and
 reminiscent of a first-order phase transition.
This large range of transformation behavior is observed despite 
relatively minor changes in the structure of our initial LDA samples.
This seemingly intractable degree of complexity can be brought within a single 
framework when we consider the potential 
energy landscape (PEL) properties sampled by the system 
during these LDA-HDA transformations.  We show that when the initial LDA
sample is located deeper within the LDA megabasin of the PEL, then the LDA-HDA transformation is
 reminiscent of a first-order phase transition.  
At least for the samples we consider, we show that the
energy of the deepest PEL local minimum
sampled by LDA during compression  quantitatively 
correlates to the sharpness of the transition.
We discuss how the PEL formalism thereby 
provides a relatively simple way to understand the complex phenomenology of glassy water 
(both real and simulated), as well as the qualitative differences found in simulations
using different water models, such as ST2 and SPC/E, following identical protocols.

This work builds upon extensive computer simulations of ST2 water in the liquid
and glassy state performed over the last several years. Specifically, we draw from
Refs.~\cite{poole2005,pooleDynamics,poolePRL2011,smallen,pabloAnisimov} where the 
phase diagram of ST2 water including the LLPT are described, and simulation studies of glassy 
water using both the ST2 and SPC/E water models~\cite{chiu1,chiu2,MySciRep}. 
 In particular, Ref.~\cite{paperI} is a PEL study of ST2 water 
during the LDA-HDA transformations described in Ref.~\cite{chiu1}. 
The present work extends the ideas Ref.~\cite{paperI}, and applies them to understand the 
sensitivity of the LDA-HDA transformation to the preparation of the intial LDA sample.

The structure of this work is as follows.  In Sec.~\ref{simulSec} we discuss our computer simulation methods.  In Sec.~\ref{LDAfromLIQTo} we study the LDA-HDA transformations for LDA samples prepared via isobaric instantaneous cooling at $P=0.1$~MPa.  A study of the LDA-HDA transformation starting from LDA samples prepared by isothermal decompression of HDA samples is presented in Sec.~\ref{LDAfromHDA}.  Sec.~\ref{summary} includes a summary and discussion.

\section{Methods}
\label{simulSec}

We perform out-of-equilibrium molecular dynamics (MD)
simulations of water using the ST2 model~\cite{ST2model}, with the 
long-range electrostatic interactions treated using the reaction field
technique~\cite{reacField}. 
Our implementation of the ST2 model is identical to that described
in Refs.~\cite{poole2005,chiu1}. 
In all cases described below, we simulate $N = 1728$ water molecules in a cubic cell
with periodic boundary conditions.  Our simulations are conducted at fixed
temperature $T$ and pressure $P$, where $T$ and $P$
are controlled using a Berendsen thermostat and barostat; see Ref.~\cite{chiu1} for details.

We consider the properties of LDA samples prepared by three distinct methods.  The first 
method is a reference case previously described in Ref.~\cite{chiu1}.  These LDA configurations are prepared from a liquid system equilibrated at $P=0.1$~MPa and $T=350$~K.  This liquid state is then cooled to either $T=180$~K or $80$~K using a cooling rate of $q_c=30$~K/ns, while maintaining constant $P=0.1$~MPa.  In the following we refer to LDA samples formed by this cooling method as ``LDA-c".  This preparation method is analogous to the experimental procedure used to produce the LDA form known as hyperquenched glassy water (HGW), although we use a faster cooling rate than in experiments;  see discussions in Refs.~\cite{chiu1,chiu2,jessina}.

In the second method we use to prepare LDA samples, we start with liquid configurations
 equilibrated at various starting temperatures 
$T_0=255$, 260, 265,~\dots , 290 and $300$~K, all at $P=0.1$~MPa.
After equilibration, these liquid configurations are cooled instantaneously to  $T=80$~K.  
We refer here to these instantaneously cooled LDA samples as ``LDA-i". 
These LDA-i samples are analogous to HGW obtained using a
cooling rate $q_c \rightarrow \infty$.  It follows that our starting LDA-i samples 
have the same density and are structurally identical to the equilibrium liquid at $T_0$, 
and hence $T_0$ is a well-defined fictive 
temperature (see, e.g., Refs.~\cite{tool,narayanaswami,hodge,kobFrancescoTis,myPREold})
for each.  

Our third set of LDA samples is prepared by starting from our LDA-c samples.  
We isothermally compress the LDA-c samples obtained at $T=80$~K or $180$~K to $P=1700$~MPa, 
a pressure sufficient to transform all samples to HDA. The rate of compression is $q_P=300$~MPa/ns.
When starting from LDA-c configurations, this value
of $q_P$ leads to sharp LDA-HDA transformations, 
similar in character to those observed experimentally using much slower rates~\cite{chiu1,MySciRep,jessina}.
The HDA forms so produced are then isothermally decompressed (also at $q_P=300$~MPa/ns) back to the LDA state at various pressures in the range $P_0 < 0$~MPa.  
We refer to these LDA samples recovered by decompression from HDA as ``LDA-d".

As described below, we subject the LDA samples formed by these three methods to compression and decompression in order to observe the sharpness of the resulting transformations between LDA and HDA.  In all cases presented here, the compression and decompression rate used is $q_P=300$~MPa/ns.  
Unless indicated otherwise, in order to characterize the compression/decompression behavior of the LDA form produced by each of our three methods, we carry out $10$ runs starting from independently generated LDA samples, to account for the sample-to-sample variation in the non-equilibrium state. 
When averaging over these 10 runs, the error bars in our plots 
represent one standard deviation of the mean.

The procedure to study the PEL of our system during 
the compression/decompression of LDA and HDA is identical to that followed in Ref.~\cite{paperI}, to which we refer the reader for details.
Briefly, during the compression/decompression runs, configurations are saved
every $10$~MPa.  For each configuration, the structure of the system at the nearest local minimum of the PEL, commonly called the inherent structure (IS),  
is obtained using the conjugate gradient algorithm~\cite{conjGrad}.
The energy of the system at this local minimum is the IS energy $E_{IS}$.
The virial expression for the pressure at the IS configuration defines 
the IS pressure $P_{IS}$,
and the curvature of the basin in the PEL at the IS is
quantified by the shape function ${\cal S}_{IS}$.
As discussed in Ref.~\cite{paperI}, the PEL properties $E_{IS}$, $P_{IS}$,
 and ${\cal S}_{IS}$ are fundamental quantities in the PEL 
formalism~\cite{francescoReview}.   
For example, for a low-temperature liquid in equilibrium (or metastable
equilibrium), the energy  and pressure of the system at a given volume $V$ and 
temperature $T$ can be 
written solely in terms of $E_{IS}$, ${\cal S}_{IS}$, and $P_{IS}$.


\section{LDA-${\rm \bf i}$}
\label{LDAfromLIQTo}

We first study the transformation behavior of our LDA-i samples, formed 
by instantaneous cooling of the equilibrium liquid from different starting temperatures $T_0$.  In this section, we conduct all compression and decompression runs at $T=80$~K.

\begin{figure}[ht]      
\centerline  { 
\includegraphics[width=7.2cm]{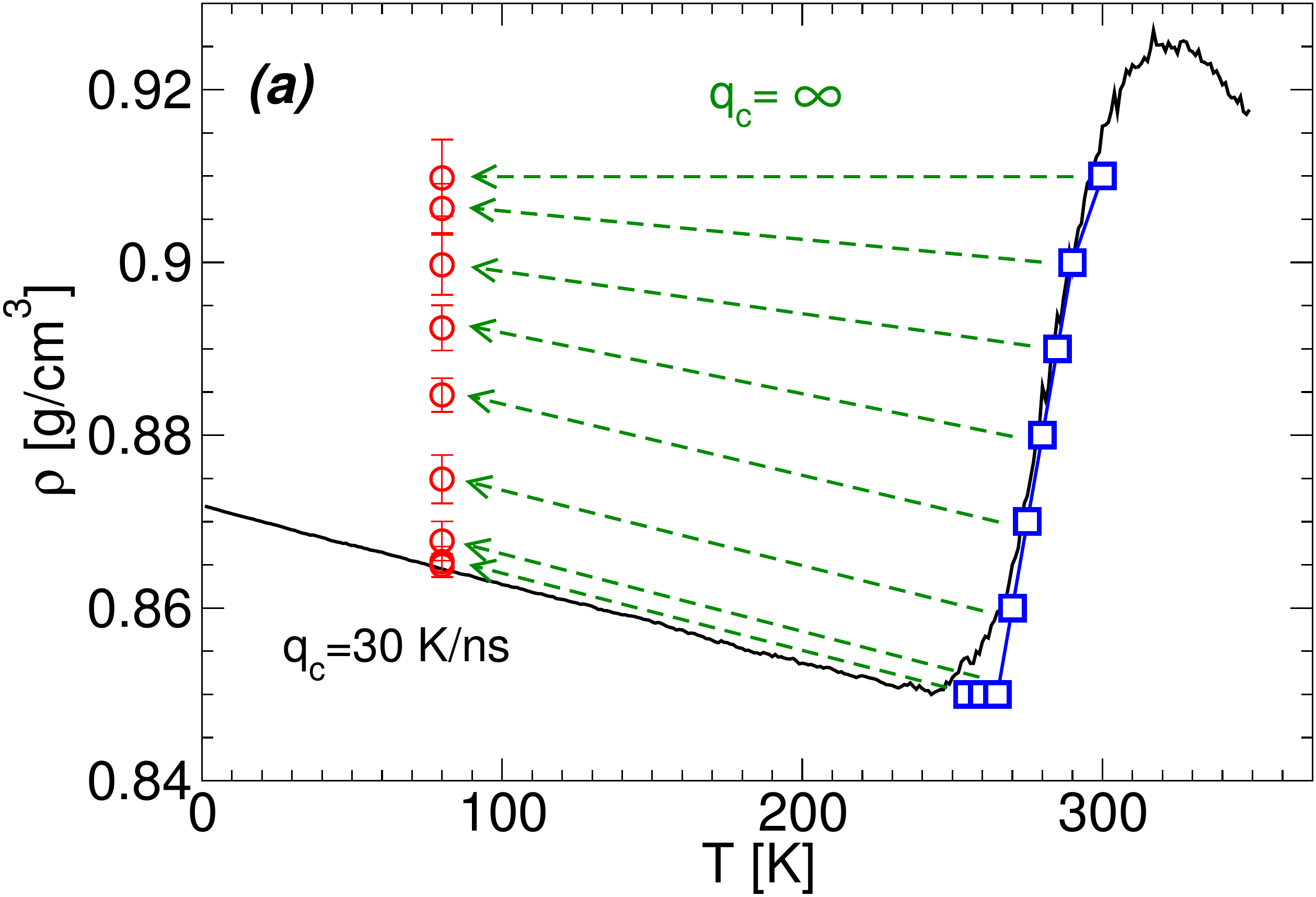}  
}
\centerline  {  
\includegraphics[width=7.0cm]{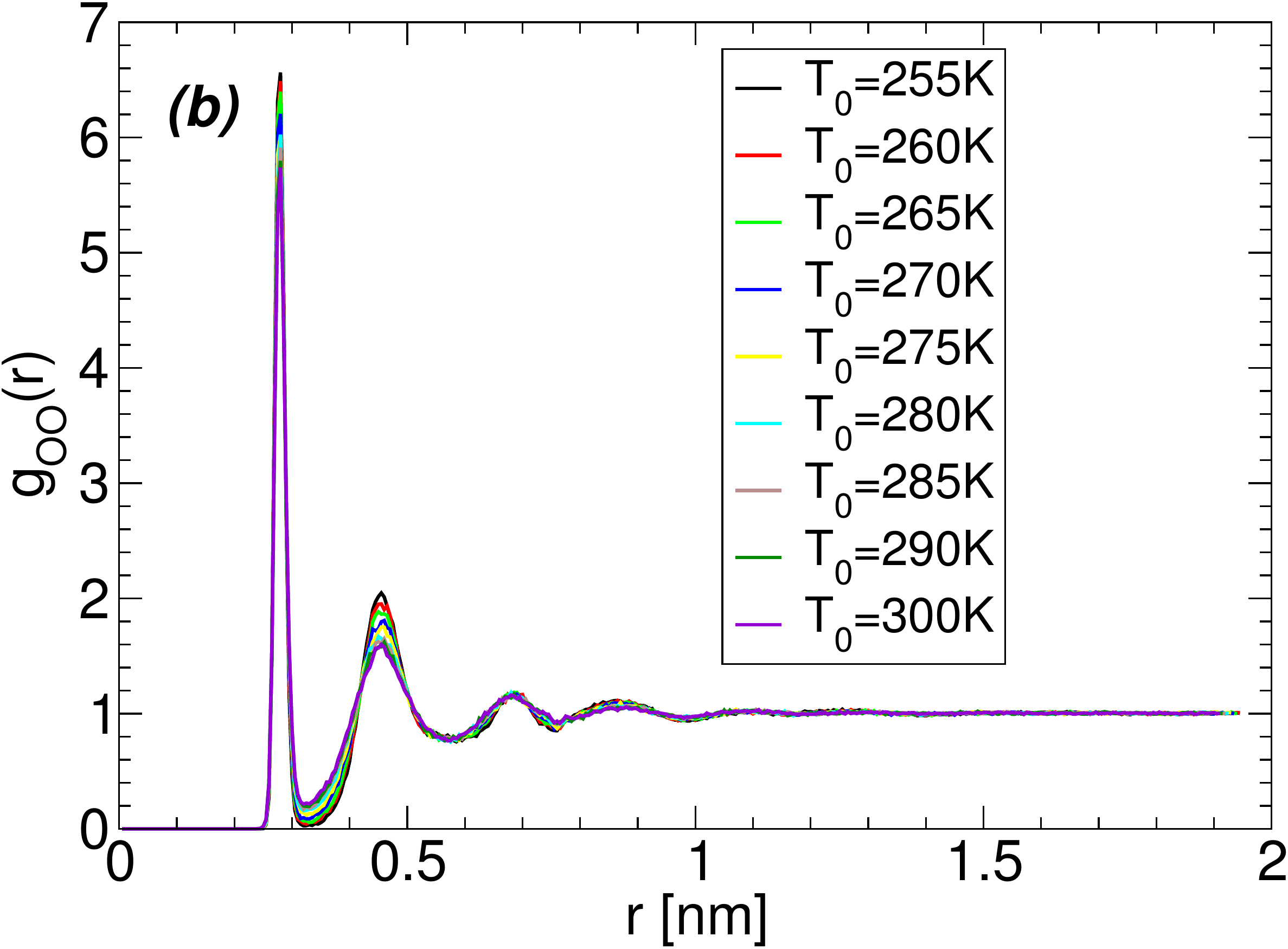}  
}
\centerline  {
\includegraphics[width=7.0cm]{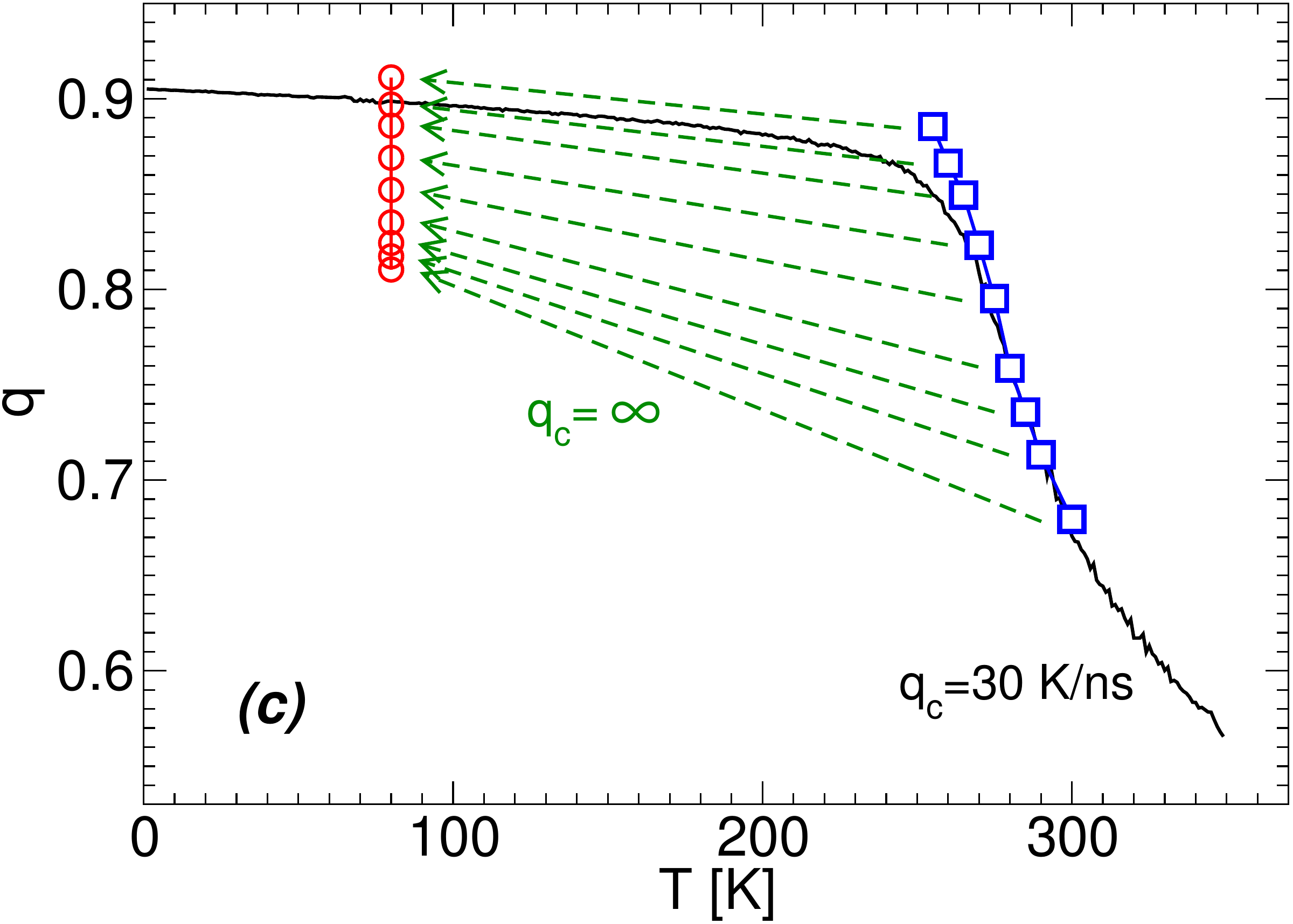}  
}
\caption{(a) Density $\rho$ of LDA-i samples (red circles) at $T=80$~K and $P=40$~MPa.
LDA-i samples are obtained from
liquid configurations equilibrated at $T_0=255,~265,\dots,290,~300$~K (blue squares) by instantaneous cooling ($q_c=\infty$) at $P=0.1$~MPa.  Green dotted lines connect each liquid state to the corresponding quenched amorphous solid state.
The black solid line shows the path by which the LDA-c sample is formed:  The liquid is equilibrated at $350$~K and then cooled at a rate of $q_c=30$~K/ns at $P=0.1$~MPa.
(b) Oxygen-oxygen RDF of our LDA-i samples 
at $T=80$~K and $P=40$~MPa [red circles in (a)]. 
(c) Average value of the tetrahedral order parameter $q$
for the LDA-i states shown in (a).
Data in all panels are an average over $10$ independent MD simulations. 
}
\label{rho-To}
\end{figure}

To characterize the properties of our LDA-i samples as a function of $T_0$, we first compress each to a (relatively low) common pressure of approximately $40$ MPa, to allow for an initial relaxation of the structure formed immediately after each quench.   
Fig.~\ref{rho-To}(a) shows the average density $\rho$ of these relaxed LDA-i samples for different values of $T_0$.  For comparison, the density of the equilibrium liquid at each value of $T_0$ is also shown.  The change in $\rho$ between the relaxed LDA-i samples and the corresponding liquid for a given $T_0$ indicates that soon after the compression starts,  
the LDA-i samples become denser than the corresponding parent liquid.
This effect becomes more pronounced as $T_0$ decreases. 
The change in $\rho$ during this initial relaxation of our LDA-i samples
is consistent with the increase in $\rho$ that occurs during the isobaric cooling process by which LDA-c is formed when the liquid is well out of equilibrium for $T<240$~K, also shown in Fig.~\ref{rho-To}(a).

The oxygen-oxygen radial distribution function (RDF) of each relaxed LDA-i sample is shown in Fig.~\ref{rho-To}(b) for various $T_0$.
These RDFs are rather similar to each other, and are 
consistent with the experimentally determined structure of LDA~\cite{chiu1}.
The effect of decreasing $T_0$ is to increase the height of the first two peaks and the
depth of the first minimum of the RDF.  That is,
as $T_0$ decreases, the LDA-i form becomes more structured.  The region between the first and second coordination shells
becomes less populated, and almost empty for $T_0=255$~K.  
This behavior suggests that LDA-i becomes more tetrahedral as $T_0$ decreases. 
We confirm th\color{black}is relationship by evaluating
the tetrahedral order parameter $q$ defined in Ref.~\cite{errington} for each relaxed LDA-i sample.  Fig.~\ref{rho-To}(c) shows that $q$ increases as $T_0$ decreases, as expected.

\subsection{LDA-HDA Transformations of LDA-$\rm i$ Samples}
\label{LDAHDA-To}

\begin{figure}[ht]      
\centerline  {   
\includegraphics[width=7.0cm]{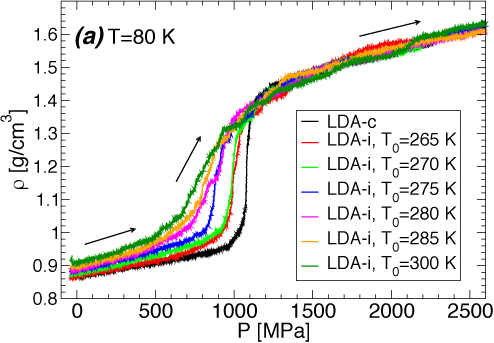}  
}
\centerline  {   
\includegraphics[width=7.0cm]{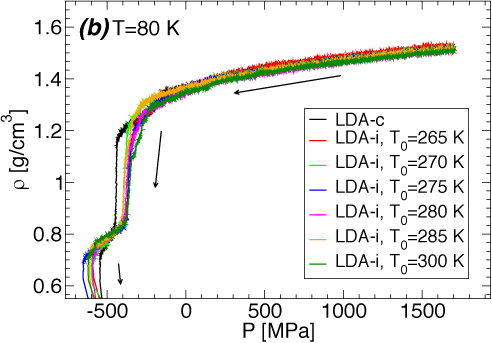}  
}
\centerline  {   
\includegraphics[width=7.0cm]{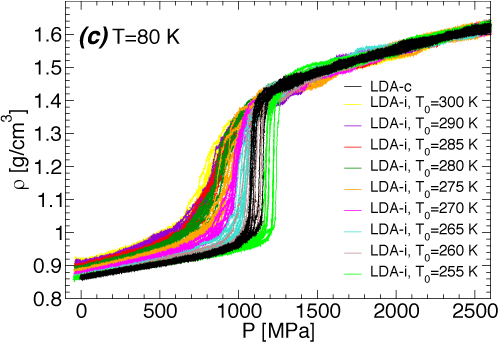}  
}
\caption{(a) $\rho(P)$ from single runs starting from LDA-i samples 
 showing the LDA-to-HDA transformation at $T=80$~K.  
For comparison, we also show the result for a single compression run
 obtained by starting from a LDA-c sample formed at $T=80$~K.
(b) Single decompression runs at $80$~K starting from the HDA configurations produced 
at $P\approx 1700$~MPa during the runs shown in (a).
{\color{black}
The large density jump at $P \approx -400$~MPa corresponds to the HDA-to-LDA 
transformation;  the density jump at $P<-500$~MPa corresponds to the LDA-to-gas transformation
(LDA fractures at these negative pressures).
}
The HDA-to-LDA transformation occurs in these runs in the vicinity of $P=-400$~MPa.
(c) $\rho(P)$ for the 10 compression runs starting from LDA-c and 
LDA-i samples.
{\color{black}  Data corresponding to LDA-c (black lines in all panels) 
are taken from Ref.~\cite{chiu1}.  }
}
\label{rho-P_HGWTo}
\end{figure}

Next, we focus on the compression-induced LDA-to-HDA transformation and decompression-induced 
HDA-to-LDA transformation of LDA-i at $T=80$~K. 
Fig.~\ref{rho-P_HGWTo}(a) shows $\rho$ as a function of $P$
during {\color{black} a single compression run}, starting from LDA-i configurations corresponding 
to different values of $T_0$; {\color{black} the behavior of $\rho(P)$
 for all ten runs, at a given temperature, are shown in Fig.~\ref{rho-P_HGWTo}(c)}.
 For comparison, we include 
$\rho(P)$ for the compression of our LDA-c samples.
As shown in 
{\color{black} Figs.~\ref{rho-P_HGWTo}(a) and \ref{rho-P_HGWTo}(c),}
 the main effect of changing $T_0$ 
is to alter the sharpness of the LDA-to-HDA 
{\color{black} transformation}.  As $T_0$ decreases, the LDA-to-HDA transformation for LDA-i 
becomes more abrupt and, at the lowest $T_0$, it becomes quite similar to the 
behavior of LDA-c.

\begin{figure}[ht]      
\centerline  {   
\includegraphics[width=7.0cm]{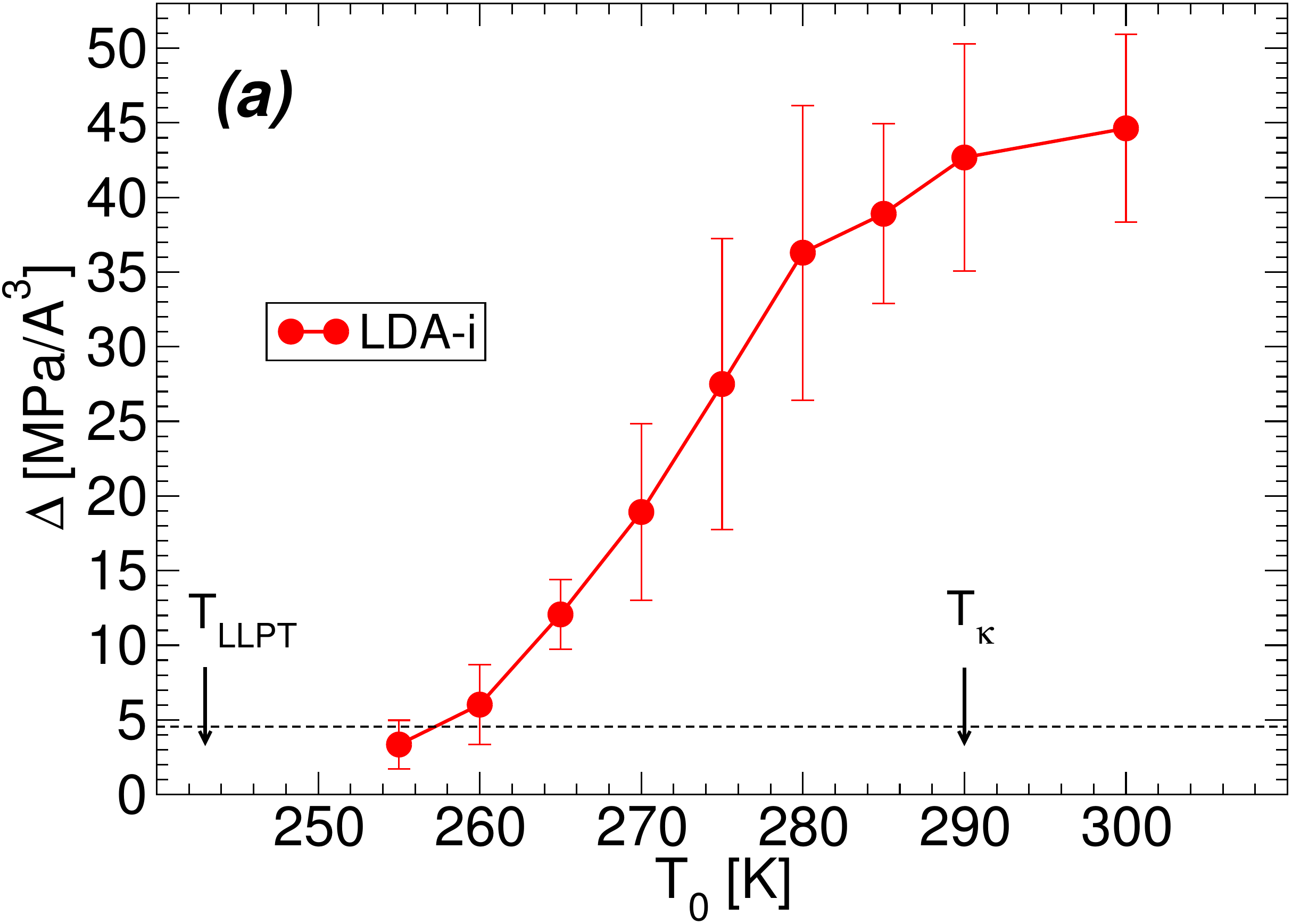}  
}
\centerline  {   
\includegraphics[width=7.0cm]{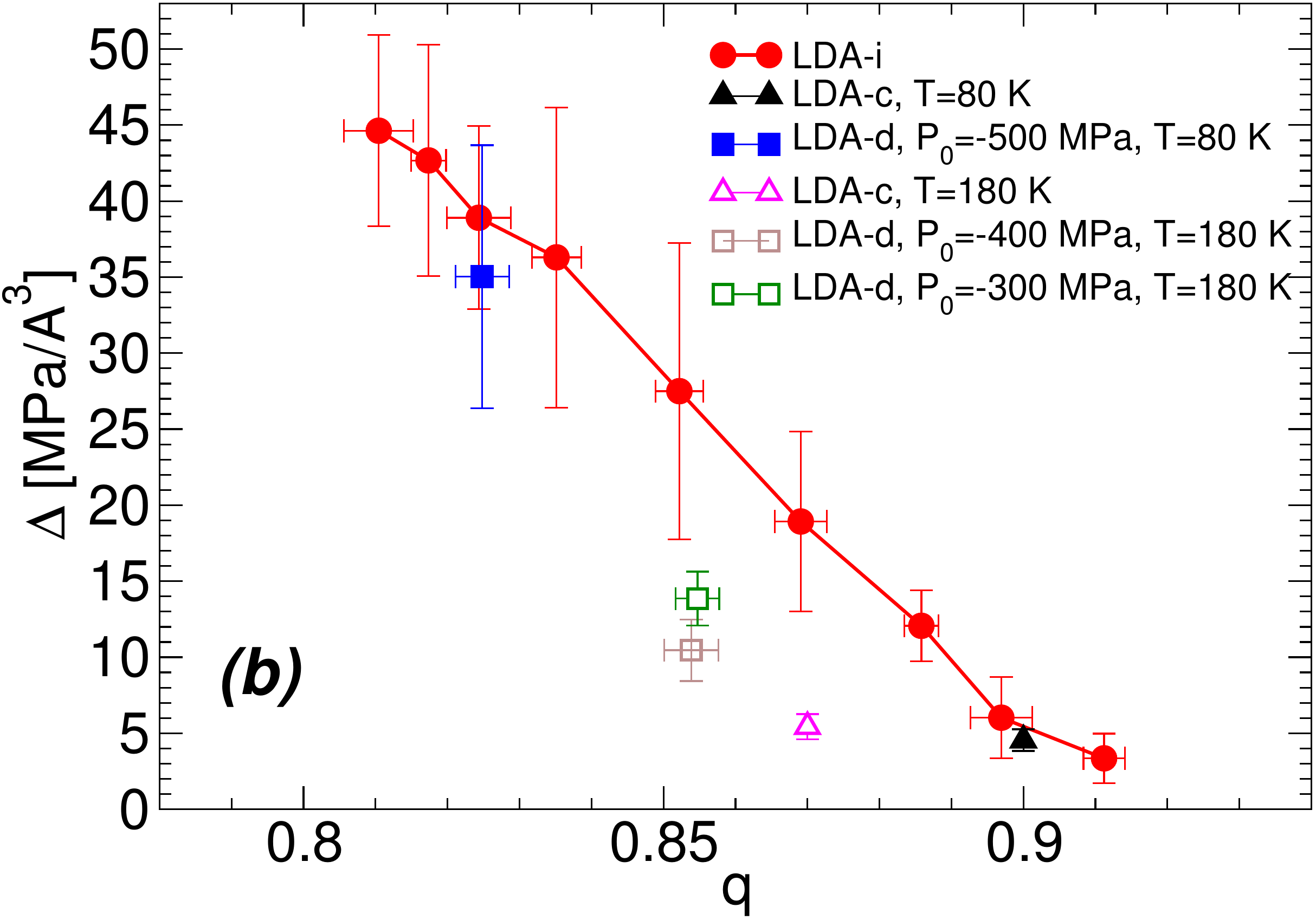}  
}
\centerline  {   
\includegraphics[width=7.0cm]{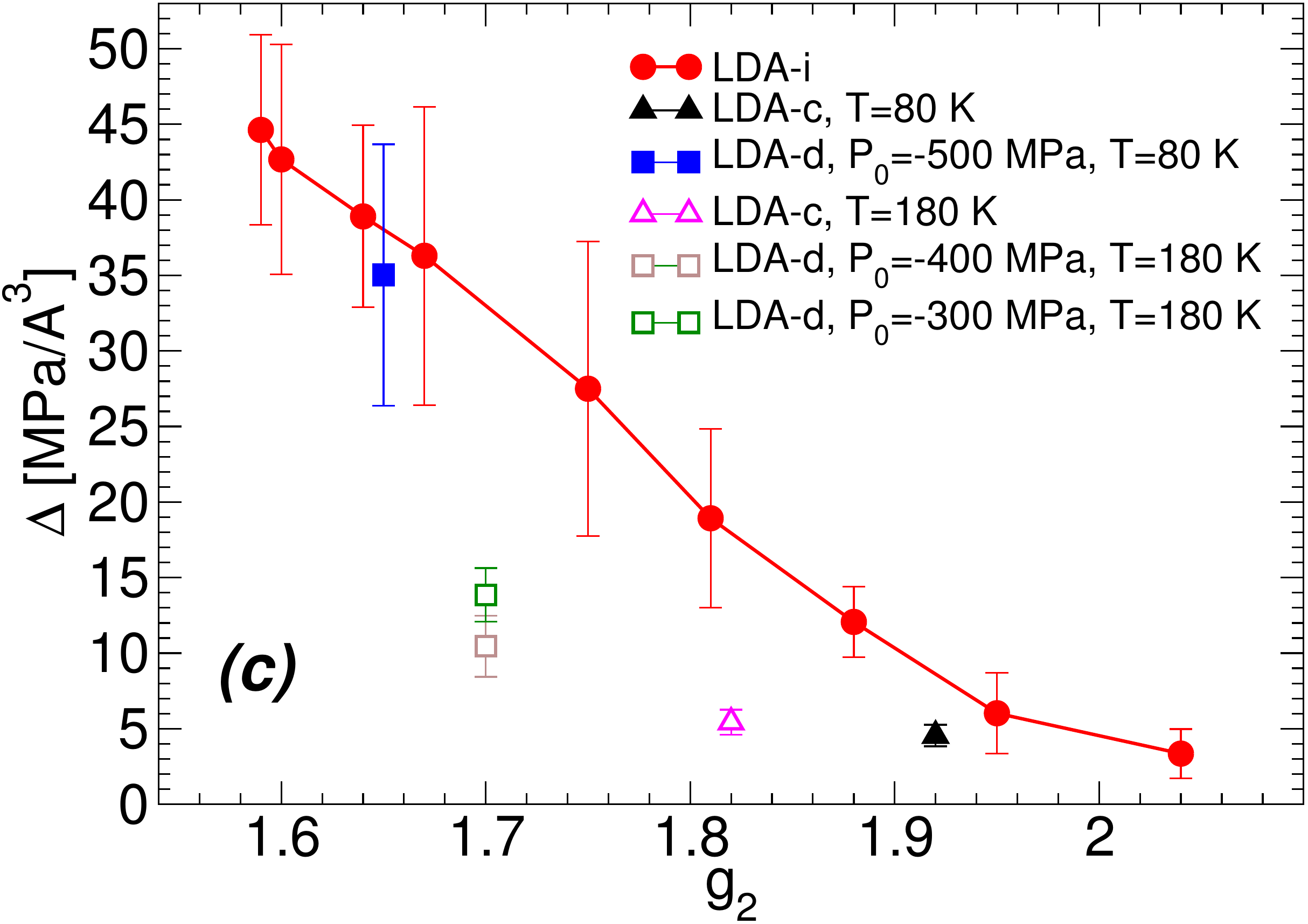}  
}
\caption{(a) $\Delta$ as a function of $T_0$ for each of our LDA-i samples (red circles).  The horizontal dotted line gives the value of $\Delta$ found for the LDA-c sample at $80$~K~\cite{chiu1}.
The vertical arrows indicate the temperature $T_{\rm LLPT}$ of the critical point of the LLPT~\cite{poolePRL2011,poole2005}
and the temperature $T_\kappa$ of compressibility maximum  at $P=0.1$~MPa~\cite{poole2005}.
(b) $\Delta$ as a function of the tetrahedral order parameter $q$ for our 
LDA-i samples at $80$~K and $40$~MPa (filled red circles); 
the LDA-c sample at $80$~K (filled black triangle); 
the LDA-c sample at $180$~K (open magenta triangle); 
the LDA-d sample at $80$~K and $-500$~MPa, recompressed to $\rho_{\rm min}$ (filled blue square);
the LDA-d sample at $180$~K and $-300$~MPa, recompressed to $\rho_{\rm min}$ (open green square);
the LDA-d sample at $180$~K and $-400$~MPa, recompressed to $\rho_{\rm min}$ (open brown square).
(c) $\Delta$ as a function of $g_2$ for the same samples presented in (b).
}
\label{dVdP_HGWTo}
\end{figure}

To quantify the sharpness of the LDA-to-HDA transformations, we calculate  
the average value of the slope $\Delta = - \left( \partial P/ \partial v \right)_T$ at the
 midpoint of the density jump during the transformations shown in 
{\color{black} Fig.~\ref{rho-P_HGWTo}(c);}
 in this 
expression $v=V/N$.  The sharper the transformation is, the smaller is $\Delta$, 
so that a discontinuous change of volume corresponds to $\Delta=0$.  As shown in Fig.~\ref{dVdP_HGWTo}(a), the sharpness of the LDA-to-HDA transformation for LDA-i varies by approximately one order of magnitude 
over the range of $T_0$ examined here. 
Also shown in Fig.~\ref{dVdP_HGWTo}(a) is $\Delta$ for our LDA-c samples, which is comparable to the values found for LDA-i for the lowest $T_0$. The $T_0$ value at which $\Delta$ for LDA-i and LDA-c coincide at $T=80$~K is consistent with the $T$ range in which the liquid falls out of equilibrium during the cooling process used to prepare LDA-c
[see Fig.~\ref{rho-To}(a)].

Fig.~\ref{dVdP_HGWTo}(b) shows $\Delta$ as function of the initial values of 
{\color{black} the tetrahedral order parameter $q$} found for the LDA-i samples at 
$T=80$~K and $P=40$~MPa [i.e. the values of $q$ plotted in 
Fig.~\ref{rho-To}(c)].  Fig.~\ref{dVdP_HGWTo}(b) shows that the more tetrahedral the starting LDA-i sample is, the sharper the LDA-to-HDA transformation becomes.
While it is not evident how to determine $q$ directly in experiments, it has been noted~\cite{nilsonNature} that the height of the second maximum of the RDF, 
$g_2$, is a useful estimator of the tetrahedrality.
{\color{black} (Alternative estimators of the system's tetrahedrality, based on the OO RDF, can be defined~\cite{pathak}).}
  We show in Fig.~\ref{dVdP_HGWTo}(c) the behavior of $\Delta$ as a function of $g_2$ for our LDA-i samples, as evaluated from the RDFs shown in Fig.~\ref{rho-To}(b).  As expected, $\Delta$ shows similar behavior when plotted as a function of either $q$ or $g_2$.

In summary, the above results show that the LDA-HDA transformation becomes sharper as the structure of the initial LDA sample approaches that of a perfect random tetrahedral network (RTN).  At our lowest values of $T_0$, $\Delta$ 
approaches zero (corresponding to an infinitely sharp transformation) and $q$ approaches unity (corresponding to a perfect RTN).  Consistent with previous results obtained using the ST2 model, our simulations are thus able to access the LDA structures that produce very sharp LDA-HDA transformations that are reminicient of a first-order phase transition.  In ST2 at ambient $P$, this regime corresponds to LDA samples formed from equilibrium liquid states at $T\lesssim 260$~K.  In the $T$-$P$ phase diagram of ST2 water~\cite{poolePRL2011,poole2005}, the point at $T=260$~K and ambient $P$ is well below the temperature of the compressibility maximum (a proxy for the Widom line~\cite{limeiWidomLine,sergeyPRL2014}), and thus is deep in the region of the phase diagram where the RTN-like structure of the LDL phase dominates the properties of the liquid state, and the amorphous solids formed from it.
Note that the temperature $T=260$~K is above the liquid-liquid critical point temperature, 
$\approx 245$~K.

Our results also highlight the sensitivity of the sharpness of the LDA-HDA transformation to small changes in the RDF of the initial LDA sample.  Fig.~\ref{rho-To}(b) shows that depopulating the space between the first and second coordination shells is critical for a LDA sample to exhibit a sudden and dramatic collapse of its hydrogen bond network upon compression, giving a sharp LDA-HDA transformation.  Although all of our LDA-i samples have RDFs consistent with the LDA family of low-density glasses, relatively small variations in their internal structure have a large influence on the sharpness of the LDA-HDA transformation observed when they are compressed.
   
\begin{figure}[ht]      
\centerline  {   
\includegraphics[width=7.0cm]{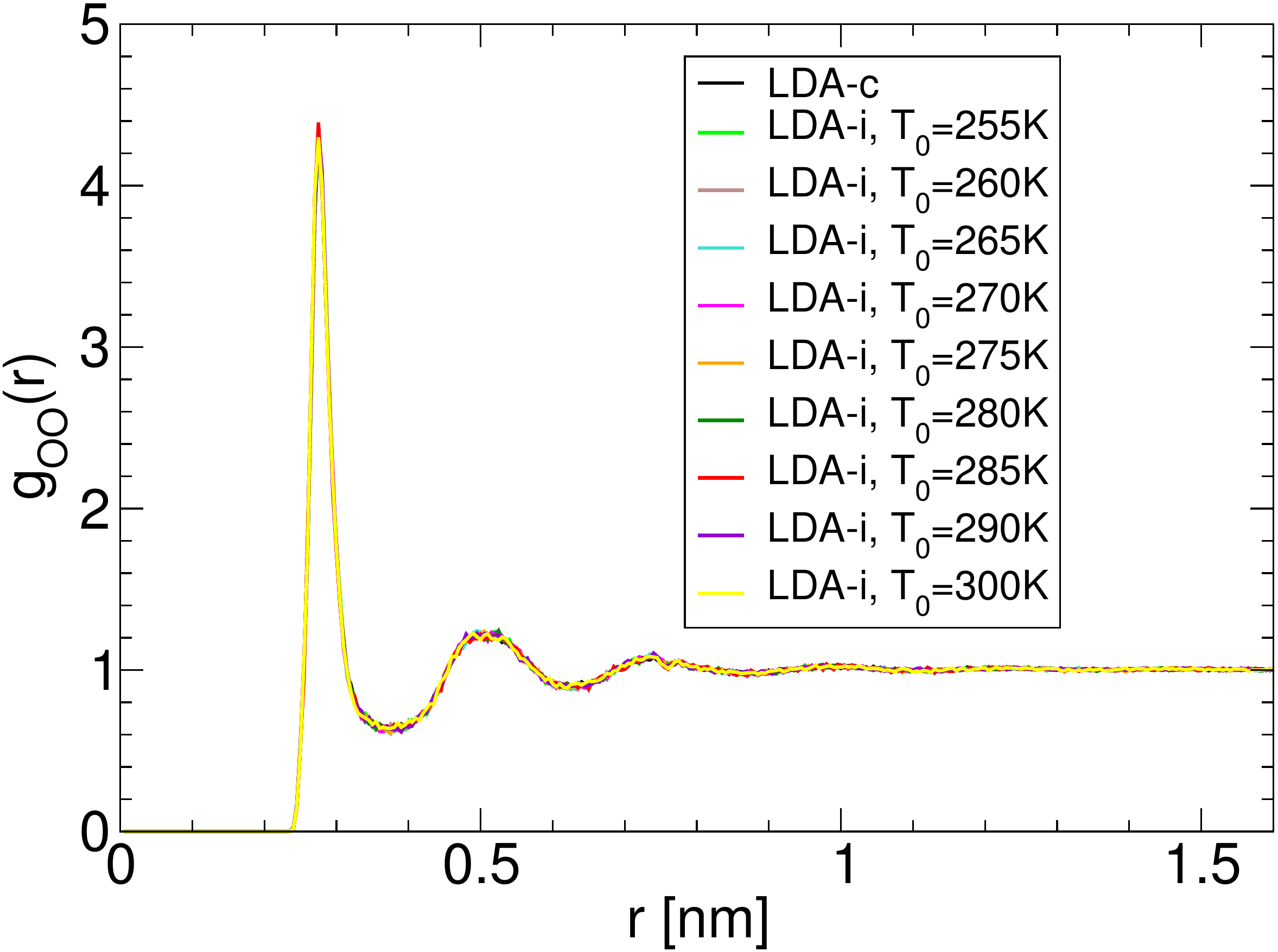}  
}
\caption{Oxygen-oxygen RDFs of the HDA configurations obtained 
at $P \approx 1700$~MPa by compression of LDA-i samples for various $T_0$.  For comparison, we also show the RDF for HDA obtained at $P=1610$~MPa by compression of the LDA-c sample at $80$~K.
}
\label{RDFs_HDA_To}
\end{figure}

Finally, we test if the properties of the HDA formed by compression of LDA-i samples depends
 on $T_0$.  Fig.~\ref{RDFs_HDA_To} shows the RDFs for HDA at $P\approx 1700$~MPa for 
each value of $T_0$.  These RDFs are indistinguishable within the noise of the data.  
Fig.~\ref{rho-P_HGWTo}(b) shows the behavior of $\rho(P)$ during the decompression 
of each of these HDA forms as a function of $T_0$.  In all cases, there is a 
relatively sharp transformation back to a LDA-like state at $P\approx-400$~MPa. 
Compared to the behavior found in Fig.~\ref{rho-P_HGWTo}(a), the 
slope of the HDA-to-LDA transformation is only weakly dependent on $T_0$. 
Together, these results suggest that once HDA forms, little `memory' remains of the initial LDA sample.

\subsection{PEL Analysis of LDA-i Samples}
\label{PEL_To}

\begin{figure}[ht]      
\centerline{   
\includegraphics[width=7.0cm]{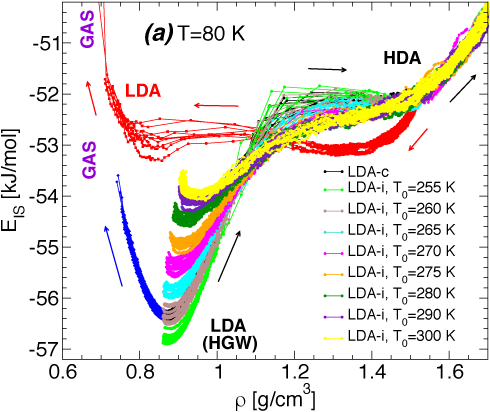}  
}
\centerline{   
\includegraphics[width=7.0cm]{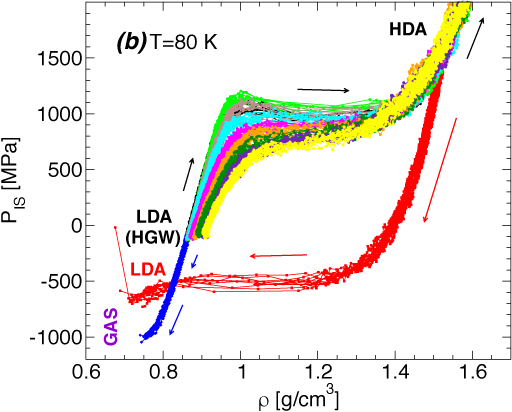}  
}
\centerline{   
\includegraphics[width=7.0cm]{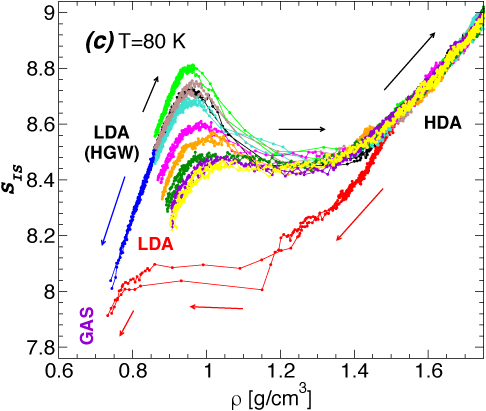}  
}
\caption{(a) $E_{IS}$, (b) $P_{IS}$, and (c)
 ${\cal S}_{IS}$  as function of $\rho$ during individual compression runs of 
the LDA-to-HDA transformations for the LDA-i and LDA-c samples shown in
 Fig.~\ref{rho-P_HGWTo}(c). 
In the case of LDA-c we also include the
decompression-induced HDA-to-LDA transformation (red lines),
as well as the decompression of the original LDA-c sample from $P=0.1$~MPa (blue lines). 
{\color{black} Data corresponding to LDA-c (black, red, and blue lines in all panels) are 
taken from Ref.~\cite{paperI}.   }
}
\label{PEL_HGWTo}
\end{figure}

To provide a single variable description that predicts the sharpness of the LDA-HDA 
transformation, we turn to the PEL properties.  
Ref.~\cite{paperI} discusses in detail the PEL behavior of the LDA-c samples.  
This behavior is reproduced in Fig.~\ref{PEL_HGWTo}.  
The LDA-c configurations are compressed isothermally at $T=80$~K, producing a sample of HDA.  This HDA form is then decompressed at the same $T$, leading to a recovered LDA sample.  The decompression process is performed until the recovered LDA sample fractures at negative pressures.  The initial LDA-c configurations obtained at $P=0.1$~MPa are also decompressed to negative pressure until they fracture.  Fig.~\ref{PEL_HGWTo} shows the behavior of $E_{IS}$, $P_{IS}$, and ${\cal S}_{IS}$ for LDA-c during this compression/decompression cycle.   

Ref.~\cite{paperI} demonstrated that three phase-transition-like phenomena are observed in the PEL properties when LDA-c is compressed through the LDA-HDA transformation:
(i) a van der Waals-like loop in $P_{IS}$; (ii) negative curvature in $E_{IS}$ as a function of $\rho$; and (iii) non-monotonic variation of ${\cal S}_{IS}$ with $\rho$.
In the same study, similar changes in $E_{IS}$, $P_{IS}$, and ${\cal S}_{IS}$ were observed during the first-order phase transition that occurs in the ST2 model when the liquid phase converts to ice VII under compression.  Ref.~\cite{paperI} therefore established the phase-transition-like character of the LDA-HDA transformation of ST2 water when examined in terms of the PEL.  Ref.~\cite{paperI} also presented evidence that the PEL for the ST2 model consists of two broad megabasins, separated by a potential energy barrier.  One megabasin corresponds to LDA and LDL configurations, and the other to HDA and HDL configurations.

Fig.~\ref{PEL_HGWTo} 
shows $E_{IS}$, $P_{IS}$, and ${\cal S}_{IS}$
during the compression of our LDA-i samples corresponding to different values of 
$T_0$.  At the starting density ($\rho=0.83$ to $0.85$~g/cm$^3$) the system is in the 
LDA megabasin, while at high-density ($\rho=1.3$ to $1.4$~g/cm$^3$)
the system is in the HDA megabasin.
We see from Fig.~\ref{PEL_HGWTo}(a) that the main effect of reducing $T_0$ 
is to bring the starting LDA-i samples deeper into the LDA megabasin. 
Moreover, it follows from Fig.~\ref{PEL_HGWTo}(c) that the individual basins explored in the LDA megabasin are ``narrower" (i.e.
they have larger curvature ${\cal S}_{IS}$) as $T_0$ decreases.
Interestingly, we note that the compression behavior shown in Fig.~\ref{PEL_HGWTo}  for the LDA-i sample for $T_0=260$~K follows almost exactly the behavior observed for the LDA-c sample, suggesting that these two forms of LDA are equivalent glasses, despite their different preparation histories. 

Combining the results of Fig.~\ref{PEL_HGWTo}(a)-(c), we find that when an LDA sample is prepared that lies deeper in the LDA megabasin, the more pronounced is the phase-transition-like character of the LDA-HDA transformation in the PEL, as quantified by the behaviors (i), (ii) and (iii) listed above.  The transformation itself [see Fig.~\ref{dVdP_HGWTo}(a)] is also sharper.
To quantify this relationship, we first note that the minima of $E_{IS}$ in 
Fig.~\ref{PEL_HGWTo}(a) 
associated with the LDA megabasin all occur in the vicinity of $\rho_{\rm min}=0.9$~g/cm$^3$.  
In order to compare configurations at a common density within the LDA megabasin, we define $E_{IS}^{\rm min}$ as the value of $E_{IS}$ for a given sample as it is compressed through $\rho=\rho_{\rm min}$.  
The relationship between the sharpness of the LDA-HDA transformation and the depth reached by the initial LDA sample in the LDA megabasin is shown in Fig.~\ref{Delta-Eis}, 
a parametric plot of $\Delta$ versus $E_{IS}^{\rm min}$ for each LDA-i sample with different $T_0$ values.  
The consistent trend shown in Fig.~\ref{Delta-Eis} suggests that $E_{IS}^{\rm min}$ may be a useful 
predictor of the compression behavior of the LDA glass, a point that is supported by data from other preparations
 of LDA, discussed in the following sections.  
Fig.~\ref{PEL_HGWTo} also illustrates that even though our LDA-i samples all have very similar densities near the minimum of the LDA megabasin, their compression behavior can vary widely, and that additional measures (such as $E_{IS}^{\rm min}$) are required to predict the behavior of a given LDA sample.

Our results also provide a framework for interpreting the LDA-HDA transformation observed 
using other computer simulation models.  For example, a study of the LDA-HDA
 transformation in SPC/E water found that the phase-transition-like behaviors of the 
PEL [properties (i), (ii), and (iii) listed above] were absent or barely observable~\cite{yoSPCE-LDAHDA}.  
No LLPT has been observed in the supercooled liquid phase of SPC/E water. 
 In Ref.~\cite{paperI}, it was proposed that the less dramatic character of 
the LDA-HDA transformation in SPC/E model arises 
because the LDA samples prepared for the compression and decompression runs were quenched from 
liquid states well above the temperature of any LLPT that might occur in this model.  The present 
results show that, even for ST2, a model that exhibits a clear LLPT, if the initial LDA samples are 
prepared with a fictive temperature $T_0$ that is well above the temperature of the LLPT,
 then the LDA-HDA transformation will lose its phase-transition-like characteristics, 
both in the directly measured thermodynamic properties (Fig.~\ref{dVdP_HGWTo}) and in the PEL 
(Fig.~\ref{PEL_HGWTo}).

\begin{figure}[ht]      
\centerline{   
\includegraphics[width=7.0cm]{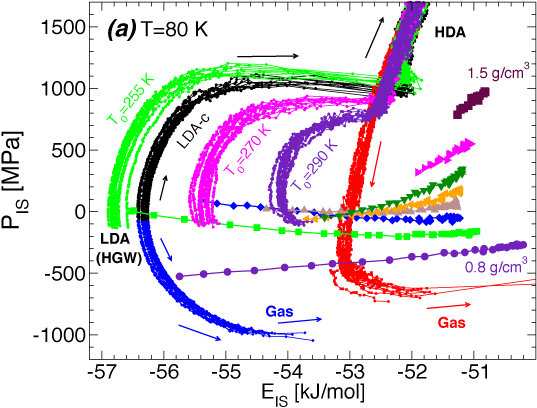}  
}
\centerline{   
\includegraphics[width=7.0cm]{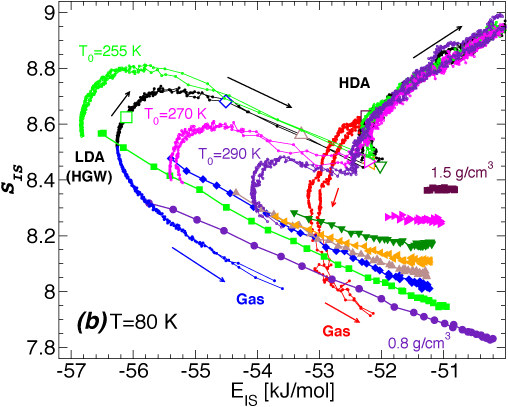}  
}
\caption{Parametric plots of (a) $P_{IS}(E_{IS})$ and (b) ${\cal S}_{IS}(E_{IS})$ 
for LDA-c, and for LDA-i samples
at various $T_0$, based on the data shown in Fig.~\ref{PEL_HGWTo}.
Also included are $P_{IS}(E_{IS})$ and ${\cal S}_{IS}(E_{IS})$ for the equilibrium liquid
at $\rho=0.8,~0.9,~1.0,...,1.5$~g/cm$^3$.
}
\label{PisSisEis_HGWTo}
\end{figure}

Finally, we note that Ref.~\cite{paperI} compared the IS visited by the LDA-c samples during the LDA-HDA transformation, with the IS explored by the equilibrium liquid at different $\rho$. 
It was found that in ST2 water, the regions of the PEL sampled
by the liquid (LDL and HDL) and the glass (LDA and HDA) differ.  Similar results have been reported for the case of SPC/E water.
Here we show that the same conclusion applies to the LDA-HDA transformations observed for all our LDA-i samples.
Figs.~\ref{PisSisEis_HGWTo}
shows  $P_{IS}(E_{IS})$ and ${\cal S}_{IS}(E_{IS})$ for the LDA-i samples corresponding to
selected values of $T_0$, and for the equilibrated liquid at different $T$.
For comparison, we include the results from Ref.~\cite{paperI} for LDA-c.
In all cases, the IS sampled by the system during 
the LDA-HDA transformation depart from the IS sampled by the liquid soon after the compression starts.
We also note that all LDA samples transform to a HDA form having identical values of
 $E_{IS}$, $P_{IS}$, and ${\cal S}_{IS}$, again suggesting that all LDA forms transform to the 
same HDA state.   

\section{LDA-$\rm \bf d$}
\label{LDAfromHDA}

In this section, we analyze the behavior of our LDA-d samples.  Our goal is to examine the 
behavior of LDA-like glasses that have not been generated directly from equilibrium liquid state configurations,
 and therefore have no well-defined fictive temperature.  Despite this, 
we will see that the compression behavior of 
our LDA-d samples can be understood in common with our LDA-i samples, using the properties of the PEL.

\subsection{LDA-HDA Transformations of LDA-d Samples}
\label{rho-ReComp}

\begin{figure}[ht]	
\centerline  {   
\includegraphics[width=7.0cm]{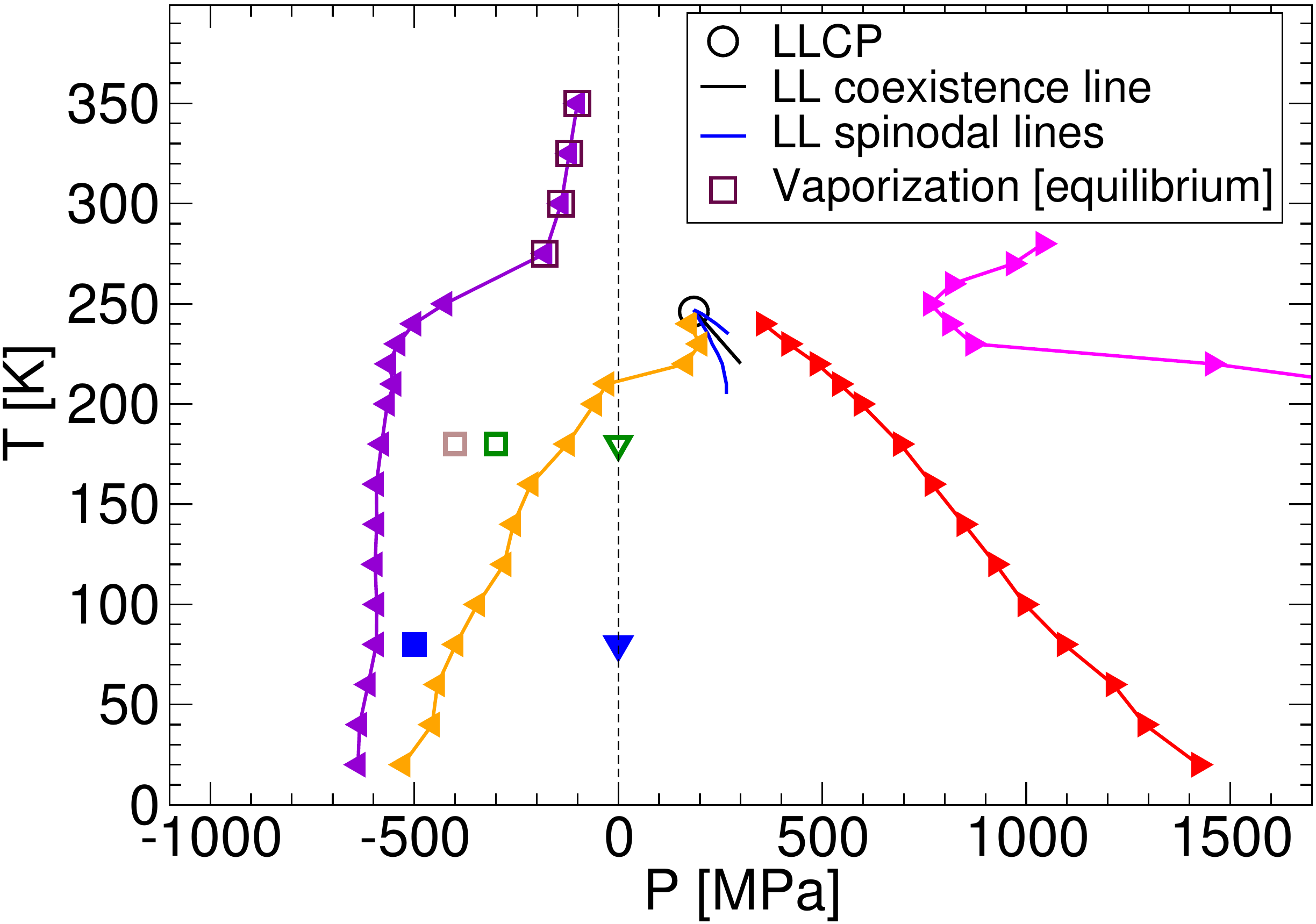}  
}
\caption{Phase diagram of glassy ST2 water based on
 isothermal compressions of LDL/LDA-c and decompressions of HDL/HDA. Adapted from
[J. Chiu, F. W. Starr, and N Giovambattista, J. Chem. Phys. {\bf 139}, 184504 (2013)],
 with the permission of AIP Publishing.
Compressions and decompressions are performed at a rate $q_P=300$~MPa/ns. 
Red (orange) triangles are the pressure-induced LDL/LDA-to-HDL/HDA (HDL/HDA-to-LDL/LDA) transformation pressures;
magenta triangles indicate the lowest pressure at which crystallization to ice VII is observed during compression.
Violet triangles are the pressure at which recovered LDA fractures; this line merges smoothly with the 
liquid-to-gas {\color{black} spinodal} line (maroon squares).  
Blue and black lines are, respectively, the spinodal and coexistence lines of the LLPT; the circle locates the critical point of the LLPT.
Blue, brown, and green symbols locate the recovered LDA-d and HDA-d samples used
 here for recompression studies. Solid blue (empty green) down-triangle indicates
 HDA-d recovered at $P=0.1$~MPa and $T=80$~K ($T=180$~K).  
The blue solid square indicates recovered LDA-d at $T=80$~K and $P_0=-500$~MPa;
brown and green empty squares locate LDA-d at $T=180$~K and $P=-400,~-300$~MPa.
}
\label{PT-states}
\end{figure}

We consider LDA-d samples prepared at both $T=80$ and $180$~K.  Starting from LDA-c samples at
 these two $T$, we compress the system to $P=1700$~MPa, and then decompress to $P_0=-500$~MPa at $80$~K; 
and to $P_0=-300$ and $-400$~MPa at $180$~K.  As shown in 
Fig.~\ref{PT-states},
this procedure brings these systems back 
to the LDA state,
providing three distinct LDA-d samples at the state points identified by the squares at $T=80,~180$~K
in Fig.~\ref{PT-states}. 
We note that within accessible simulation time scales, the systems at both $T=80$ and $180$~K show
 no liquid-like relaxation, and hence can be considered to be in the glass state.
This is consistent with the temperature dependence of the equilibrium relaxation time, 
the extrapolation of which vastly exceeds our simulation times scales at the temperatures 
considered. Specifically, the mode coupling temperature of ST2 water at which the relaxation 
time appears to diverge is $T_{MCT}=270$~K at $P= 0.1$~MPa~\cite{pooleDynamics}.

\begin{figure}[ht]      
\centerline  {
\includegraphics[width=7.0cm]{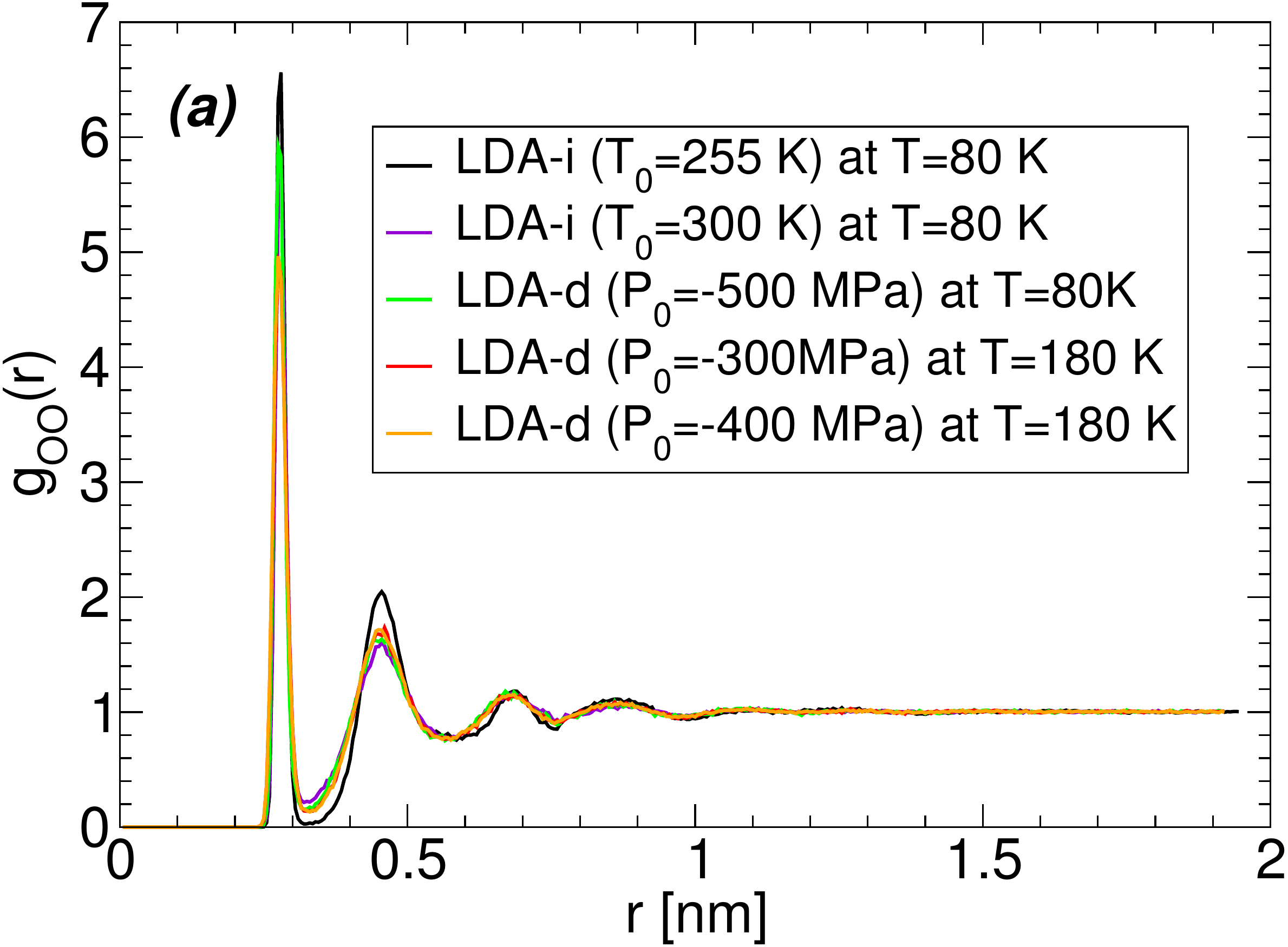}  
}
\centerline  {
\includegraphics[width=7.0cm]{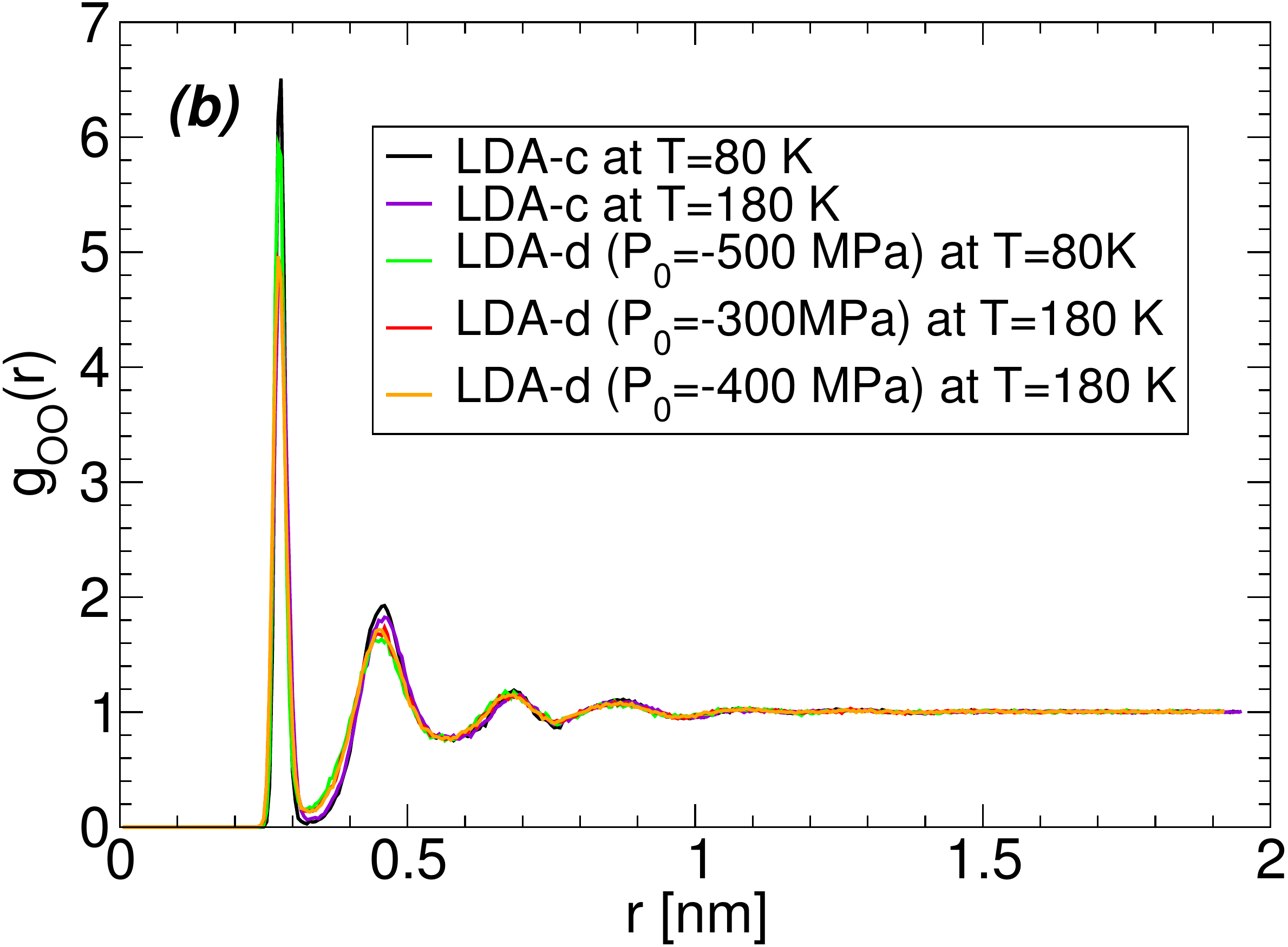}  
}
\caption{Oxygen-oxygen RDFs of our three LDA-d samples, each recompressed to $\rho_{\rm min}$.
{\color{black} (a)  Comparison of the RDF of LDA-d and 
 LDA-i for the cases $T_0=255$ and $300$~K [from Fig.~\ref{rho-To}(b)].
(b) Comparison of the RDF of LDA-d and LDA-c.  }
}
\label{RDF-LDAd}
\end{figure}

{\color{black} We recompress our three LDA-d samples until their densities are all close 
to $\rho_{\rm min}$.  This recompression to $\rho_{\rm min}$ allows us to compare the 
structure of these samples all at the same density.  As we have seen in the previous 
section, $\rho_{\rm min}$ seems to be the relevant system density to consider because
 the properties at this density provide a way to predict the behavior of the 
sample when compressed; this possibility is explored further below.  }
The RDFs of these LDA-d samples are compared to LDA-i 
{\color{black} and LDA-c in Fig.~\ref{RDF-LDAd}(a) and (b).  }
We also create two forms of HDA recovered at $P_0=0.1$~MPa, before the HDA-to-LDA transformation 
occurs during the decompression process;  see the down triangles in Fig.~\ref{PT-states}.  We refer to these samples as ``HDA-d".

\begin{figure}[ht]      
\centerline  {  
\includegraphics[width=9.0cm]{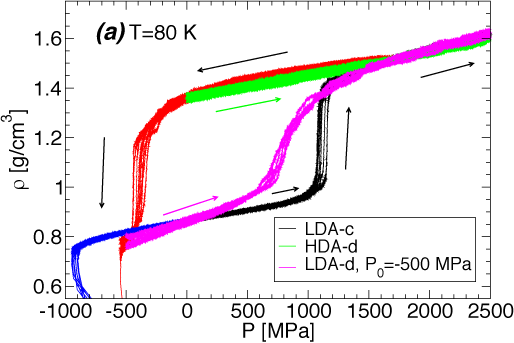}  
}
\centerline  {  
\includegraphics[width=9.0cm]{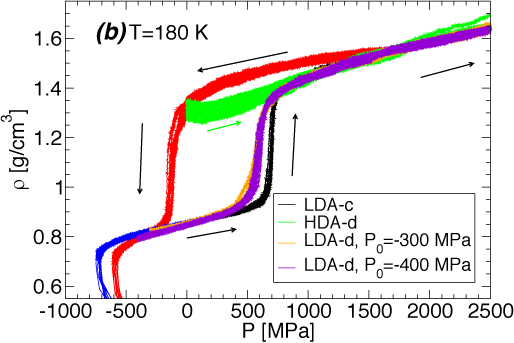}  
}
\caption{(a) $\rho$ as function of $P$ during the recompression of our LDA-d (magenta lines) 
and HDA-d samples (green lines) at $T=80$~K (solid blue square and down-triangle 
in Fig.~\ref{PT-states}). 
(b) Same as in (a) for recovered LDA-d 
($P_0=-400,~-300$~MPa; violet and orange lines, respectively) and
HDA-d (green lines) at $T=180$~K (empty brown and green squares, and empty green down-triangle in Fig.~\ref{PT-states}). 
For comparison, we include in (a) and (b) $\rho(P)$ obtained 
during the compression of LDA-c (black lines), the decompression
of the resulting HDA form (red lines), and the decompression of the 
original LDA-c sample starting from $P=0.1$~MPa (blue lines).
}
\label{recompLDA-T}
\end{figure}

We then recompress our LDA-d and HDA-d samples at the same $T$ ($80$ or $180$~K) at which they were prepared.
The behavior of $\rho(P)$ during these recompression runs is shown in Fig.~\ref{recompLDA-T}.  For comparison, we include $\rho(P)$ for the LDA-c sample for which the LDA-HDA transformation is especially sharp. 
At both $T=80$ and $180$~K, $\rho(P)$ during
recompression of the HDA-d samples follows closely the decompression path by which they were formed.
For example, the green and red lines in Fig.~\ref{recompLDA-T}(a) almost overlap, 
suggesting a reversible compression/decompression process for HDA in the range $P=0.1$ to $1700$~MPa at $T=80$~K. 
 At $T=180$~K,
some differences occur between the the green and red lines in Fig.~\ref{recompLDA-T}(b). 
At this temperature, the HDA-d sample is very close to the HDA-to-LDA transformation 
line (the orange boundary in Fig.~\ref{PT-states}) and hence, some evolution in $\rho(P)$ during 
recompression is not surprising.

When the LDA-d samples are recompressed at their respective $T$, none show a LDA-HDA transformation as sharp as we observe for LDA-c.  The LDA-HDA transformation at $T=80$~K is especially gradual, while the transformations observed at $180$~K are closer in sharpness to the LDA-c case, although still not as sharp.  The sharpness of these three LDA-d transformations, as quantified by $\Delta$, is shown in Fig.~\ref{dVdP_HGWTo} as a function of both $q$ and $g_2$ for the LDA-d samples.  As in Fig.~\ref{RDF-LDAd}, we have evaluated $q$ and $g_2$ for the LDA-d samples after recompressing them to $\rho_{\rm min}$, to bring each sample to a common density near the minimum of the LDA megabasin.
Within the error of our calculations, the correlation of $\Delta$ with both $q$ and $g_2$ for 
the LDA-d sample at $80$~K is consistent with the trend found for our LDA-i samples at the 
same $T$.  However, the data clearly do not collapse to a single functional form, indicating that 
$q$ and $g_2$ do not uniquely predict the sharpness of the LDA-HDA transition.

\subsection{PEL Analysis of LDA-d Samples}
\label{PEL-ReComp}

\begin{figure}[ht]      
\centerline{  
\includegraphics[width=7.0cm]{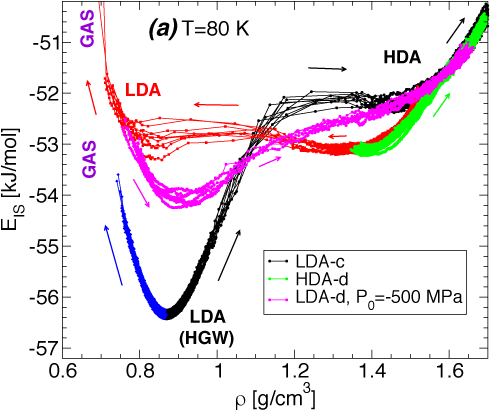}  
}
\centerline{  
\includegraphics[width=7.0cm]{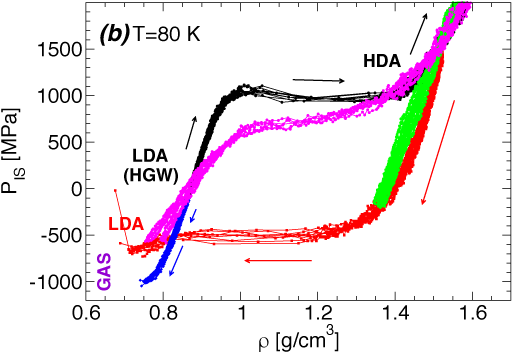}  
}
\centerline{  
\includegraphics[width=7.0cm]{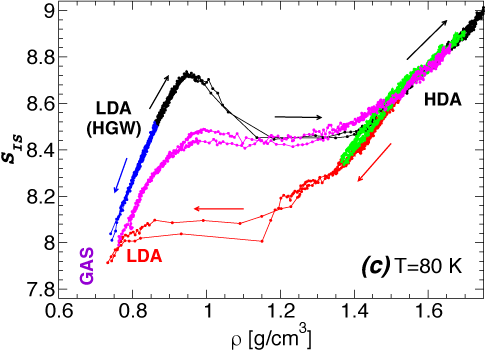}  
}
\caption{(a) $E_{IS}$, (b) $P_{IS}$, and (c) ${\cal S}_{IS}$  as function of $\rho$ 
for the compression and decompression runs shown in 
Fig.~\ref{recompLDA-T}(a).  Data is labeled using the
same colors as in Fig.~\ref{recompLDA-T}(a).
{\color{black} Data corresponding to LDA-c (black, red, and blue lines in all panels) are
taken from Ref.~\cite{paperI}.   }
}
\label{PEL-rho_reCompPo-T80}
\end{figure}

\begin{figure}[ht]      
\centerline{  
\includegraphics[width=7.0cm]{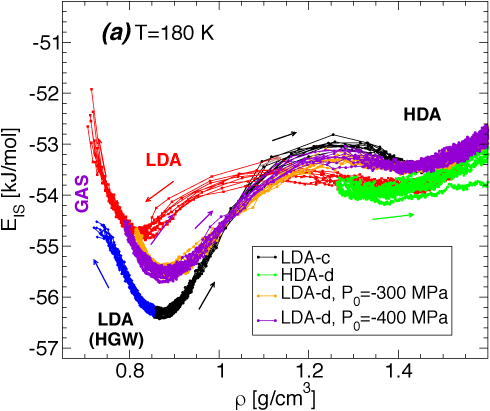}  
}
\centerline{  
\includegraphics[width=7.0cm]{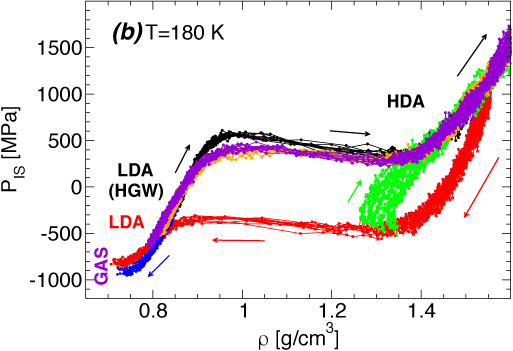}  
}
\centerline{  
\includegraphics[width=7.0cm]{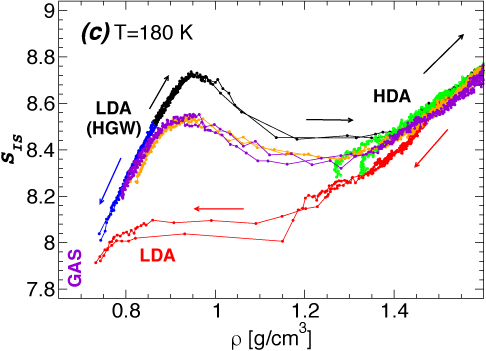}  
}
\caption{
(a) $E_{IS}$, (b) $P_{IS}$, and (c) ${\cal S}_{IS}$  as function of $\rho$ 
for the compression and decompression runs shown in 
Fig.~\ref{recompLDA-T}(b).  Data is labeled using the
same colors as in Fig.~\ref{recompLDA-T}(b).
}
\label{PEL-rho_reCompPo-T180}
\end{figure}

Figs.~\ref{PEL-rho_reCompPo-T80} and 
\ref{PEL-rho_reCompPo-T180}
show the variaton of 
$E_{IS}$, $P_{IS}$, and ${\cal S}_{IS}$ with $\rho$ during the 
recompression of our three LDA-d samples.
In Figs.~\ref{PEL-rho_reCompPo-T80} and \ref{PEL-rho_reCompPo-T180} we include for reference the PEL properties sampled by our LDA-c sample.
We also show $E_{IS}$, $P_{IS}$, and ${\cal S}_{IS}$ during the recompression of our two HDA-d samples.

Fig.~\ref{PEL-rho_reCompPo-T80}(a)  shows that the LDA-d sample at $P_0=-500$~MPa 
and $T=80$~K starts its recompression path at a value of $E_{IS}$ quite far above that of LDA-c.  During recompression, $E_{IS}$ for this LDA-d sample passes through a minimum very similar to the minimum explored by the LDA-i samples with the highest values of $T_0$.  Consistent with this similarity, the sharpness of the LDA-HDA transformation for this LDA-d sample 
is low (see Fig.~\ref{dVdP_HGWTo}), and the phase-transition-like behavior in $P_{IS}$
 and $S_{IS}$ [Figs.~\ref{PEL-rho_reCompPo-T80}(b) and (c)] is absent or very weak.  These 
results suggest that our LDA-d sample at $P_0=-500$~MPa is a rather poorly structured 
configuration within the LDA megabasin, with a correspondingly weak transition 
from LDA to HDA upon compression.

In Fig.\ref{PEL-rho_reCompPo-T180}, we show the corresponding plots for the LDA-d samples prepared at $P_0=-300$ and $-400$~MPa, for $T=180$~K.  Here $E_{IS}$ for the initial LDA-d samples starts out closer to the LDA-c curve in Fig.\ref{PEL-rho_reCompPo-T180}(a), and both $P_{IS}$ and $S_{IS}$ [Figs.~\ref{PEL-rho_reCompPo-T180}(b) and (c)] display more robust signatures of phase-transition-like behavior in the PEL.  
As shown in Fig.~\ref{dVdP_HGWTo}, the values of $\Delta$ 
for LDA-d and LDA-c samples at $180$~K are closer to each other than at 
$80$~K.

Taken together, Figs.~\ref{PEL-rho_reCompPo-T80} and \ref{PEL-rho_reCompPo-T180} demonstrate that the PEL properties of a given sample of LDA correlate well to the sharpness of the LDA-HDA transformation observed upon compression of these samples, regardless of the details of the path by which the samples are prepared.  Our initial LDA-d samples are stressed amorphous solids located relatively high in the PEL of the LDA megabasin.  While they recover somewhat as they are compressed, passing through a minimum in $E_{IS}$, this restructuring is not sufficient to allow exploration of the deepest regions of the LDA megabasin, which are better represented by the LDA-c samples.  Our LDA-d samples thus illustrate that poorly structured LDA ice will display a poorly defined LDA-HDA transition.  
To confirm this interpretation, we have evaluated $\Delta(E_{IS}^{\rm min})$ for each of our LDA-d samples, as shown in Fig.~\ref{Delta-Eis}.  Remarkably, Fig.~\ref{Delta-Eis} shows that
the values of $\Delta$ for LDA-d show the same
$E_{IS}^{\rm min}$-dependence
in approximately the same way as all our other LDA samples,
 within the limits of our uncertainty.  
This finding indicates that $E_{IS}^{\rm min}$ provides a one-to-one mapping 
(for a given compression rate) to predict
 the emergence of a genuine first-order change in density of the LDA-HDA transformation, 
regardless of the sample preparation.  Consideration of this prediction for other 
models and sample preparations will be valuable to validate or refute its universality.

\begin{figure}[ht]      
\centerline{
\includegraphics[width=8.0cm]{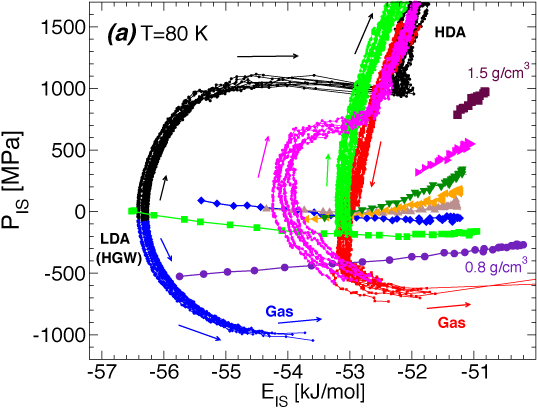}  
}
\centerline{
\includegraphics[width=8.0cm]{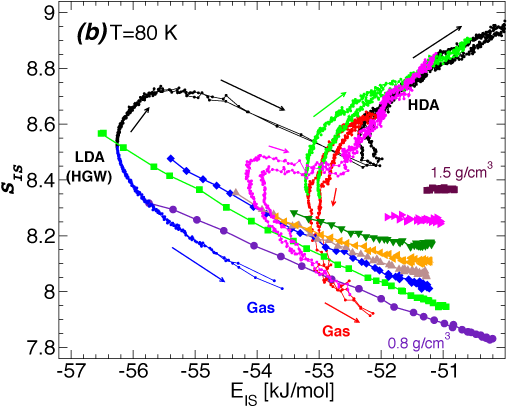}  
}
\caption{
Parametric plots of (a) $P_{IS}(E_{IS})$ and (b) ${\cal S}_{IS}(E_{IS})$ 
for the data shown in Fig.~\ref{PEL-rho_reCompPo-T80} at $T=80$~K.  Data is shown with same colors as in Fig.~\ref{PEL-rho_reCompPo-T80}.
Also included are $P_{IS}(E_{IS})$ and ${\cal S}_{IS}(E_{IS})$ for the equilibrium liquid
at $\rho=0.8,~0.9,~1.0,...,1.5$~g/cm$^3$.  
}
\label{PisEis_reCompPo-T80K}
\end{figure}
\begin{figure}[ht]      
\centerline{
\includegraphics[width=8.0cm]{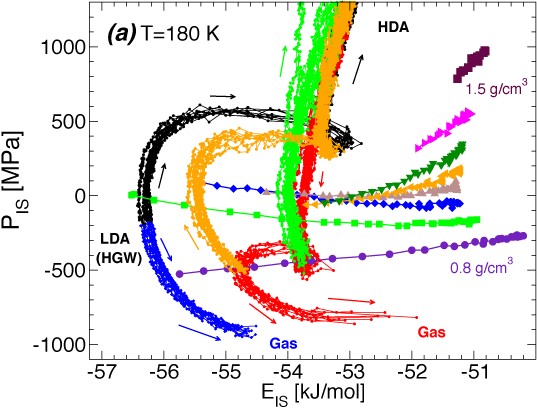}  
}
\centerline{
\includegraphics[width=8.0cm]{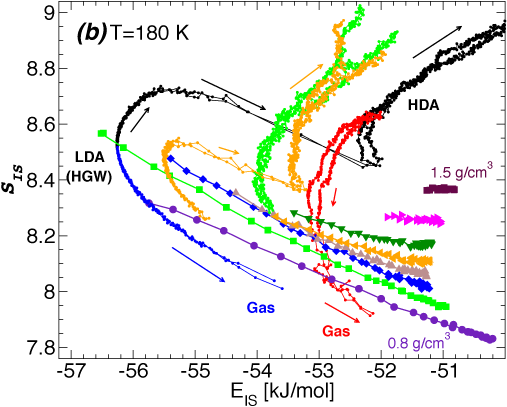}  
}
\caption{
Parametric plots of (a) $P_{IS}(E_{IS})$ and (b) ${\cal S}_{IS}(E_{IS})$ 
for the data shown in Fig.~\ref{PEL-rho_reCompPo-T180} at $T=180$~K.  Data is shown with same colors as in Fig.~\ref{PEL-rho_reCompPo-T180}.
For clarity, we have omitted the results during recompression of LDA-d recovered at $P_0=-400$~MPa. 
Included are $P_{IS}(E_{IS})$ and ${\cal S}_{IS}(E_{IS})$ for the equilibrium liquid
at $\rho=0.8,~0.9,~1.0,...,1.5$~g/cm$^3$.  
}
\label{PisEis_reCompPo-T180K}
\end{figure}

We conclude this section by comparing the IS sampled by the equilibriium liquid 
and our LDA-d samples during compression.  Figs.~\ref{PisEis_reCompPo-T80K} and \ref{PisEis_reCompPo-T180K} 
show the evolution of  
$P_{IS}(E_{IS})$ and ${\cal S}_{IS}(E_{IS})$ during the recompression of LDA-d, along with
the corresponding values for the equilibrium liquid over a range of $\rho$.  For comparison, we include in each figure $P_{IS}(E_{IS})$ and ${\cal S}_{IS}(E_{IS})$ for LDA-c.
Again we observe that during the compression of LDA-d at the present rates, the system explores IS never sampled by the equilibrium liquid.  The present results, taken together with those of Refs.~\cite{yoSPCE-LDAHDA} and \cite{paperI} suggest that this difference is a general feature of all LDA forms, regardless of the preparation procedure, at least for the compression rates accessible in MD simulations.

\section{Summary and Discussion}
\label{summary}

\begin{figure}[ht]      
\centerline{
\includegraphics[width=8.0cm]{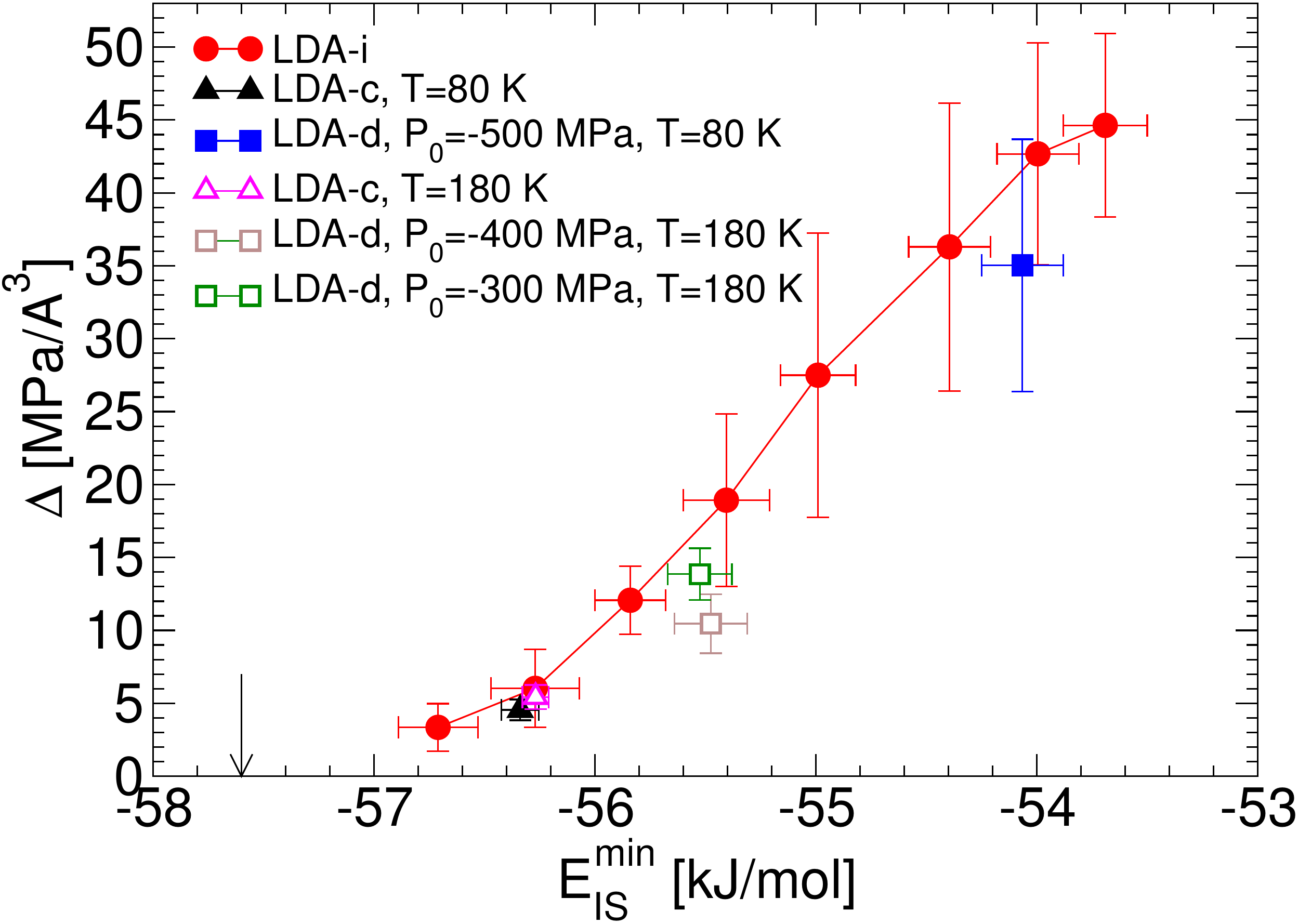}  
}
\caption{$\Delta$ as function of $E_{IS}^{\rm min}$ for all LDA samples considered here.  Symbols are the same as in Fig.~\ref{dVdP_HGWTo}(b). The vertical arrow indicates $E_{IS}=-57.6$~kJ/mol, the estimated value for a perfect RTN in ST2 water~\cite{pooleDynamics}.
}
\label{Delta-Eis}
\end{figure}

In summary, we have examined the LDA-to-HDA transformation starting from initial LDA samples
 prepared in three distinct ways.  Our main result is captured in Fig.~\ref{Delta-Eis}:  despite the 
differences in the histories of all of our samples, the sharpness of the resulting transformation to HDA 
can be predicted from the depth reached by the initial LDA sample as it passes through the LDA megabasin of the PEL.  Our results demonstrate that the abruptness of the LDA-HDA transformation can vary widely, even when using a model (ST2) for which a well-defined LLPT is known to occur.  The variability of the LDA-HDA transformation with sample preparation is therefore, by itself, not a basis for rejecting the occurrence of a LLPT in real water.  
Our results also show that this variability in transformation behavior can be understood in terms of
the effects of sample preparation on the properties of the PEL sampled by the system.

{\color{black} 
We also note that all of our initial LDA samples are obtained via procedures that
 begin by using liquid phase configurations.  We have not considered the case in which 
ice I$_h$ is compressed into the HDA state, and then decompressed to LDA ice,
 another process used frequently in experiments.  It would be interesting to check our 
results for this case as well, which we hope to present in a future work.
}

Regarding the relationship between $T_0$ and $\Delta$ presented in Fig.~\ref{dVdP_HGWTo}(a),
 we emphasize that our cooling and compression rates are several orders of magnitude 
faster than those used in experiments~\cite{jessina}.   It therefore remains an open question if our results will remain valid on experimental time scales.  Previous simulation work shows that variation of these rates over one order of magnitude shifts the behavior of $\rho(P)$ during the LDA-HDA transformation, but does not significantly change $\Delta$.  However, connecting results obtained from simulations of glassy systems to real glasses remains a challenge.

Although we have only studied the ST2 water model here, our results illuminate the differences
 found in previous work between the behavior of ST2 and SPC/E water.  We have shown that by preparing
 LDA samples with a fictive temperature above the $T$ range of the LLPT, the ST2 model 
exhibits a LDA-HDA transformation that is smooth and gradual, similar to that observed in SPC/E.  
Hence the lack of a sharp LDA-HDA transformation in a given water model does not exclude the
 possibility that a LLPT occurs in that model, only that the starting samples are high in
 the PEL compared to a possible LDA megabasin.  In other words, the method by which the 
initial sample of LDA is prepared must be taken into account.

Our results show that the sharpness of the LDA-HDA transformation can be predicted from 
the properties of the initial LDA sample.  Although the RDFs for our various LDA samples do
 not differ greatly in overall appearance, structural measures that are sensitive to the 
quality of the RTN such as $q$ and $g_2$ correlate well with $\Delta$, at least for initial 
LDA samples prepared at the same $T$ and subjected to the same compression rate,
 but do not provide a single parameter prediction for the sharpness of the transition.  Rather, 
$E_{IS}^{\rm min}$ 
is especially useful as a predictor for the sharpness of the LDA-HDA transformation.  
All of our initial LDA samples approximately fall on a single curve in Fig.~\ref{Delta-Eis}, 
including points obtained at different $T$.  We also note that the data in Fig.~\ref{Delta-Eis} 
are on track to reach $\Delta=0$ in the vicinity of $E_{IS}=-57.6$~kJ/mol, 
the estimated value of $E_{IS}$ for a perfect RTN of ST2 water~\cite{pooleDynamics}.
These values of $E_{IS}$ correspond to deep regions of the LDA megabasin
since they are very close to the IS energy of ice $I_h$ 
(the lowest possible value for the IS energy);
for example, $E_{IS} \approx -59$~kJ/mol for 
ice $I_h$ at $\rho=0.83$~g/cm$^3$ in the ST2 model~\cite{pooleDynamics}.

Our results thus suggest that it may be possible, at least for amorphous solid water, to identify a relatively small number of ``state variables" that would determine if two uncorrelated glasses (with different preparation histories) are the same, in the sense of whether they will behave the same when they are e.g. compressed or heated.  For example, we find that the behavior under compression of the LDA-c sample is approximately the same as the LDA-i sample with $T_0=260$~K.  From Fig.~\ref{PEL_HGWTo} we see that at a given value of $T$ and $\rho$, these two samples always have approximately the same values of 
$E_{IS}$, $P_{IS}$ and ${\cal S}_{IS}$, and as a consequence, they have the same value of 
$E_{IS}^{\rm min}$.
The same approximate correspondence occurs between our LDA-d sample at $T=80$~K and our LDA-i sample with $T_0=290$~K (compare Figs.~\ref{PEL_HGWTo} and~\ref{PEL-rho_reCompPo-T80}).
Although these cases are anecdotal, they suggest that the traditional state variables $T$ and $\rho$ need only be augmented by a few more observables to specify the state of the glass, and that the PEL quantities $E_{IS}$, $P_{IS}$, and ${\cal S}_{IS}$ are viable candidates for these additional state variables.  Our results therefore confirm that a systematic search for such state variables to describe glassy materials, even complex polyamorphic systems such as water, is worth pursuing, and that the PEL may be a useful framework within which to conduct this search.






\section*{Acknowledgments}

This project was supported, in part, by a grant of computer time from
the City University of New York High Performance Computing Center under
NSF Grants CNS-0855217, CNS-0958379 and ACI-1126113. PHP thanks NSERC
and ACEnet.  We thank Wesleyan University for computational resources.
FWS was supported by NIST Award 70NANB15H282.







\newpage

\newpage

\newpage

\newpage

\newpage

\newpage

\newpage

\newpage



\end{document}